\documentclass{article}

\PassOptionsToPackage{numbers,sort,compress,square}{natbib}

\usepackage[preprint]{tackling_climate_workshop_style}

\usepackage[utf8]{inputenc} 
\usepackage[T1]{fontenc}    
\usepackage{hyperref}       
\usepackage{url}            
\usepackage{booktabs}       
\usepackage{amsfonts}       
\usepackage{nicefrac}       
\usepackage{microtype}      
\usepackage{graphicx}       
\usepackage{todonotes}        
\usepackage{algorithm}
\usepackage{algorithmic}
\usepackage{subcaption}
\usepackage{amsmath,amssymb,amsfonts}
\usepackage{comment}
\usepackage[belowskip=-15pt,aboveskip=0pt]{caption}

\def \scale {\text{scale}}
\def \p {\partial}
\newcommand{\vect}[1]{\boldsymbol{\mathbf{#1}}}

\newcommand{\anonymize}[1]{\textit{Anonymous}}

\title{Ensembles of Neural Surrogates for Parametric Sensitivity in Ocean Modeling}

\author{%
  Yixuan Sun  \\
  Argonne National Laboratory\\
  \texttt{yixuan.sun@anl.gov} \\
  \And
  Romain Egele \\
  Oak Ridge National Laboratory \\
  \texttt{regele@ornl.gov} \\
  \AND
  Sri Hari Krishna Narayanan \\
  Argonne National Laboratory\\
  \texttt{snarayan@anl.gov} \\
  \And
  Luke Van Roekel \\
  Los Alamos National Laboratory \\
  \texttt{lvanroekel@lanl.gov} \\
  \And
  Carmelo Gonzales \\
  NVIDIA \\
  \texttt{carmelog@nvidia.com} \\
  \And
  Steven Brus \\
  Argonne National Laboratory\\
  \texttt{sbrus@anl.gov} \\
  \And
  Balu Nadiga \\
  Los Alamos National Laboratory \\
  \texttt{balu@lanl.gov} \\
  \And
  Sandeep Madireddy \\
  Argonne National Laboratory\\
  \texttt{smadireddy@anl.gov} \\
  \And
    Prasanna Balaprakash\\
  Oak Ridge National Laboratory \\
  \texttt{pbalapra@ornl.gov} \\
}

\begin{document}

\maketitle

\begin{abstract}

Accurate simulations of the oceans are crucial in understanding the Earth
system. Despite their efficiency, simulations at lower resolutions must rely
on various uncertain parameterizations to account for unresolved processes.
However, model sensitivity to parameterizations is difficult to quantify, making
it challenging to tune these parameterizations to reproduce observations. Deep
learning surrogates have shown promise for efficient computation of
the parametric sensitivities in the form of partial derivatives, but their
reliability is difficult to evaluate without ground truth derivatives. In this work, we 
leverage large-scale hyperparameter search and ensemble learning to improve both
forward predictions, autoregressive rollout, and backward adjoint sensitivity
estimation. Particularly, the ensemble method provides epistemic uncertainty of
function value predictions and their derivatives, providing improved
reliability of the neural surrogates in decision making.

\end{abstract}

\section{Introduction}
\begin{figure}[!t]
    \centering
    \includegraphics[width=\linewidth]{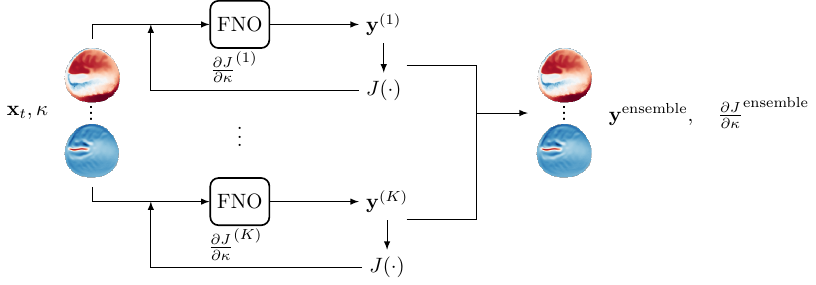}
    \caption{Overview of deep ensemble of Fourier Neural Operators (FNOs) for ocean parametric sensitivity.
    We train individual models with diverse hyperparameters to produce time-stepping 
    predictions and parametric sensitivities. We aggregate these outputs to produce ensemble
    predictions of future ocean states and estimates of the partial derivative of objective $J$ w.r.t. the parameterization $\kappa$.}
    \label{fig:overview}
\end{figure}

The ocean is a vast reservoir of heat and plays a significant role in redistributing heat from the equatorial regions to the poles.  Accurate simulation of this heat transport requires resolution of smaller scale ocean eddies~\cite{Griffies2005}.  While various ocean models have been constructed to model these complex dynamics~\cite{hoch2020mpas, golaz2022doe,
gaikwad2024mitgcmadv2opensource}, at lower resolution, these critical eddies must be modeled separately using a {\em parametrization}. Commonly, small scale eddies are modeled to remove baroclinic instability and also transport tracers along constant density layers ~\cite{GentMcwilliams1990,Redi1982}. At smaller scales, vertical turbulent mixing is modeled as a down gradient process~\cite{Large1994}.
On the larger scale ocean, the effect of parameterizations is poorly understood,
making it challenging to optimize uncertain model parameters to better match observations~\cite{perezhogin2025largeeddysimulationocean,
wolfram2015diagnosing}. Therefore, it is crucial to understand the model
parametric sensitivity to effectively perform tuning to minimize model bias relative to observations. However, the existing physics-based ocean model codes are
often too computationally expensive and not readily differentiable to support
perturbation analysis or automatic differentiation~\cite{griewank2008edp}. Alternatively, neural surrogates have emerged~\cite{kim2022review,
Radeta2022Deep, Er2023Research, chattopadhyay2024oceannet}, which can 
approximate the parametric sensitivity
inexpensively
by differentiating the trained
networks~\cite{sun2023surrogate, sun2024parametric}. 
However, without ground truth derivatives to constrain the training, such as in~\cite{olearyroseberry2023derivativeinformedneuraloperatorefficient}, the
estimated derivatives can deviate substantially from the true values, even when
the network accurately reproduces the forward process.
We propose to alleviate this issue through an ensemble of neural surrogates
constructed from a large-scale hyperparameter optimization (HPO), as
illustrated in Figure~\ref{fig:overview}. The ensemble improves both the
forward prediction performance and derivative estimates, which enables
more stable autoregressive rollouts of the ocean states for long-term forecasts.
Moreover, the neural surrogate ensemble provides quantified epistemic
uncertainty, providing detailed reliability information for decision-making.

\section{Neural Surrogate and Deep Ensemble}\label{sec:method}


The target neural network surrogate model, $M(\cdot ;\theta)$, approximates the
true physical model, $\mathcal{M}(\vect{x}, \kappa)$. Here $\vect{x}$ and
$\kappa$ are the current ocean states and physical parameter of interest, and \(\theta\) is
set of the neural network parameters (including trainable parameters and
architecture choices). We also aim to estimate the parametric sensitivity of a
model-dependent objective function $J$ by computing its derivative w.r.t.
the parameterization, $\frac{d J}{d \kappa} = \nabla_{\mathcal{M}} J
\frac{\partial \mathcal{M}}{\partial \kappa}$\footnote{The total derivative
equals the partial derivative with respect to $\kappa$ in this case, as the
current state $\vect{x}$ is assumed to be independent of $\kappa$.}.
Specifically, for given training data set $\mathcal{D}_{\rm train} = \{
(\vect{x}_i, \kappa_i), \vect{y}_i\}^N_{i=1}$, generated from the true physical
process $\vect{y}_i = \mathcal{M}(\vect{x}_i, \kappa_i)$, where $\vect{y}$
represents the ocean states at a future time, and $N$ is the number of
input-output pairs. We train the neural network, $\mathcal{N}_{\theta}$, such that $M(\vect{x}_i,
\kappa_i; \theta) = \vect{y}_i$ by minimizing the loss function $
\mathcal{L}({\theta; \mathcal{D}_{\rm train}}) = \frac{1}{N}\sum_{i=1}^N \Vert
\vect{y}^{(i)} - \mathcal{N}_{\theta}(\vect{x}^{(i)}, \kappa^{(i)})\Vert^2_2, $
and use $\frac{\partial M}{\partial \kappa}$ as the estimate of the parametric
sensitivity. 

We select the ensemble members from the results of a large-scale HPO, where the
mapping between the hyperparameters and model performance is treated as a
black-box function, $F = h(\lambda), \lambda \in \Lambda$. \(F\) is the
objective function, and \(\Lambda\) denotes the hyperparameter search space of
which a point defines the data preprocessing, neural architectures, and training
strategies. We simply use the validation loss as the search objective to be
minimized, \(F^* = \min\nolimits_{\theta} \mathcal{L}(\theta;
\mathcal{D}_{\text{val}})\).
During the search, we model the relationship using a
tree-based surrogate, \(\hat{h}\), from the current known hyperparameter set and
objective pairs, \((\lambda_i, F_i), \, \text{s.t.}\, F_i =
\hat{h}(\lambda_i)\), and perform centralized Bayesian optimization to select
the next best evaluation point that can potentially lead to the global
optimum~\cite{deephyper_software}. Once the HPO has completed, we then select
$K$ top-performing models as the ensemble members.


Let $M_k(\vect{x}, \kappa ; \theta_k)$ denote the $k$th member of the ensemble
parameterized by $\theta_k$. The ensemble prediction of the ocean states are $
\vect{y}_{ens} = \sum_{k=1}^K w_k M_k(\vect{x}, \kappa, \theta_k) $, where
\(w_k\) is the weight associated with the \(k\)th
member of the ensemble. Then the model uncertainty is expressed as the
empirical distribution variance, $ \sigma^2 = \sum_{k=1}^K w_k (\vect{y}_k -
\vect{y_{ens}})^2 $.

Regarding the parametric sensitivity, we use $J_k = \frac{1}{2} \Vert
\vect{y}_{k} \Vert_2^2$ as the scalar-valued objective\footnote{The specific
    choice of the objective is for a proof of concept, which doesn't
    necessarily correspond to a meaningful physical quantity. A more realistic
objective function can be used based on the specific application.} and compute
the individual model sensitivity estimate as $\frac{\partial J_k}{\partial
\kappa} = \frac{\partial J}{\partial \vect{y}_{k}} \frac{\partial
\vect{y}_{k}}{\partial \kappa} = \vect{y_k} \frac{\partial
\vect{y}_{k}}{\partial \kappa}$. Similar to the ensemble prediction, we
aggregate the individual model sensitivity estimates to obtain the ensemble
sensitivity estimate as $ \frac{\partial J}{\partial \kappa}^{\text{ensemble}}
= \sum_{k=1}^K w_k \vect{y_k} \frac{\partial \vect{y}_{k}}{\partial \kappa}$,
and the uncertainty of the ensemble sensitivity estimate as $ \sigma_{\partial
J}^2 = \sum_{k=1}^K w_k (\frac{\p J_k}{\p \kappa} - \frac{\p J}{\p
\kappa}^{\text{ensemble}})^2$.

\section{Numerical Experiments}\label{sec:exp}


We adopt the Fourier Neural Operator (FNO)~\cite{li2020fourier} as the surrogate
and apply our framework to the Simulating Mesoscale Ocean Activity (SOMA)
model~\cite{wolfram2015diagnosing}. The FNO is trained on five ocean state
variables, salinity, temperature, layer thickness, and meridional and zonal
velocities, using the current state and the bolus kappa parameter
(\(\kappa\) represents its strength)~\cite{GentMcwilliams1990} as input, and
predicting the next-step state. Hyperparameter optimization is performed with
DeepHyper~\cite{deephyper_software}, and the top $K=10$ models form our
ensemble. We compare two weighting schemes: uniform weights and regression-based
weights learned on the validation set. Appendix~\ref{sec:experiment} provides
details on the SOMA setup, data generation, preprocessing, normalization, and
training. Evaluation is conducted on three tasks: single-step prediction,
autoregressive rollout, and parametric sensitivity estimation.

\paragraph{Single-step prediction}

We report the model prediction RMSE over the testing set in
Table~\ref{tab:rmse}. To highlight the
effect of learning, the results also include the constant predictor, which is
the temporal mean of ocean states. Aside from the layer thickness, all trained
models present improved predictions over the constant predictor. In particular,
the optimal model is superior to the baseline in predicting four out of the five
states. The performance improvement continues for the constructed ensembles,
which achieve the lowest RMSE scores. These results showcase the proposed
framework's high accuracy in modeling the forward process of the physical model
over all single surrogates.
\begin{table}[!t]
    \caption{RMSE \((\downarrow)\) of single step forward predictions.}
    \label{tab:rmse}
    \resizebox{\linewidth}{!}{
    \begin{tabular}{cccccc}
    \toprule
         & \textbf{Constant Pred.}&\textbf{Baseline} &\textbf{Top-1}  & \textbf{Top10 Ensemble} & \textbf{Top10 Ensemble (weighted)} \\
        \midrule
    \textbf{Layer Thickness}  & \textbf{0.0004} &  0.0011  & 0.0008 & \textbf{0.0004} & \textbf{0.0004}\\
    \textbf{Zonal V.} & 0.0122 & 0.0022 &0.0024& \textbf{0.0016} & \textbf{0.0016}\\ 
    \textbf{Meridional V.} & 0.0075 & 0.0024 & 0.0019 & \textbf{0.0013} &\textbf{0.0013} \\
    \textbf{Temp.} & 0.3876 & 0.0494 & 0.0484 & \textbf{0.0385} & 0.0392\\
    \textbf{Salinity} & 0.0072 & 0.0026 & 0.0024 & 0.0018 & \textbf{0.0017} \\
    \bottomrule
    \end{tabular}}
\end{table}

\begin{figure}[!t]
    \centering
    \includegraphics[width=\linewidth]{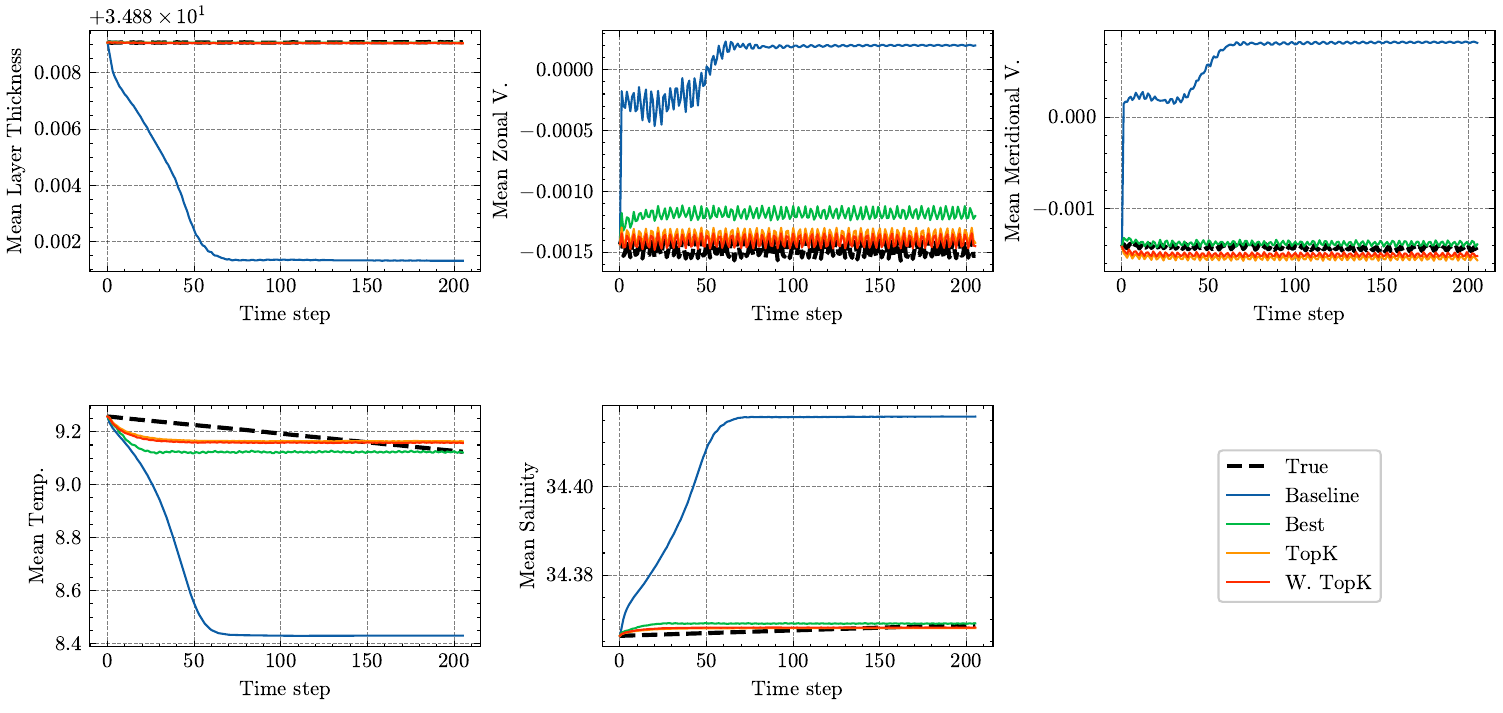}
    \captionof{figure}{Autoregressive rollout performance comparisons among the three
    models. The curves are the spatially averaged ocean states for 8 years. For
    the baseline model, the forecasts quickly diverge from the true state due to
    error accumulation, whereas the ensembles generate stable prediction for all
    ocean states.} 
    \label{fig:rollout}
\end{figure}

\paragraph{Rollout}
We investigate the model's ability for producing longer-term forecasts by
recursively making predictions of the ocean state variables at the next time step using
the previous prediction as input. In the ideal situation, when the model
prediction for single-step forward matches the truth state exactly, with the
fixed time resolution, we expect to see infinite long, accurate rollout.
However, due to the accumulation of prediction errors, the duration of
producing accurate and stable rollout shortens. To evaluate the rollout
performance, we compute the spatially averaged ocean state time series from the
true and predicted states and compare the discrepancies between the two.
Figure~\ref{fig:rollout} shows the rollout performance comparison among models.
The optimal and ensemble models show a significant rollout performance
improvement over the baseline despite the slight improvement in single
time-stepping prediction task. The long-term autogressive prediction amplifies
the subtle difference in error accumulation. 

%
%
%
%

\paragraph{Parametric sensitivity} Our final objective is to leverage the
trained models to compute parametric sensitivity in the form of derivatives. As
described in Section \ref{sec:method}, we use the simple sum-of-square of the
ocean states as the scalar-valued function for which we study the sensitivity.
We compute and visualize the parametric sensitivity for temperature fields at
\(\kappa=1429.81\). Figure~\ref{fig:sensitivity_compare} shows the comparison
of estimated sensitivity using the trained models on the sea surface (depth 0).
The models show substantial differences in magnitudes and the value
distribution in their estimate for the parametric sensitivity. The baseline and
ensemble models show similar sensitivity ranges and present value concentration
at certain areas while the best model from the HPO displays a wider range and
more uniform distribution of values. Meanwhile, the ensemble models offer
quantified uncertainties for the sensitivity estimate, suggesting evident
variations among the ensemble members. This observation validates that a
well-trained neural network for the forward model does not necessarily provide
accurate derivative estimates. As the ground truth sensitivity is not available, we instead indirectly
validate the estimated sensitivity using the linearized models to make
predictions for nearby points. We expect the model having a better sensitivity
estimation to produce a more accurate linear model. We linearize the trained
model around each $\kappa_{\rm in} \in \mathcal{D}_{\rm test}$, to obtain $
M^{\rm linearized}(\vect{x}, \kappa) = M(\vect{x}, \kappa_{\rm in}) + \frac{\p
M}{\p \kappa_{\rm in}}(\kappa - \kappa_{\rm in})$. We then use $M^{\rm
linearized}(\vect{x}, \kappa)$ to make predictions of ocean state variables under
a $\kappa$ value in the testing set such that $\kappa = \min
\{\kappa^\prime \in \mathcal{D}_{\rm test} |\kappa_{in} < \kappa^\prime\}$.
Figure~\ref{fig:linear_model} shows the testing RMSE of the linearized models
on the temperature fields, where we also report the full model performance for
reference. The baseline and uniformly weighted ensemble result in the lower
RMSE, while the linearized best model presents the highest errors, despite its
improved performance in the forward prediction (as the full model) over the
baseline. By evaluating both forward-prediction accuracy and sensitivity
estimates, our ensembles achieve superior performance while delivering
quantified predictive uncertainty. 

\vspace{.5em}
\begin{minipage}[c]{.48\textwidth}
    \centering
    \includegraphics[width=\linewidth]{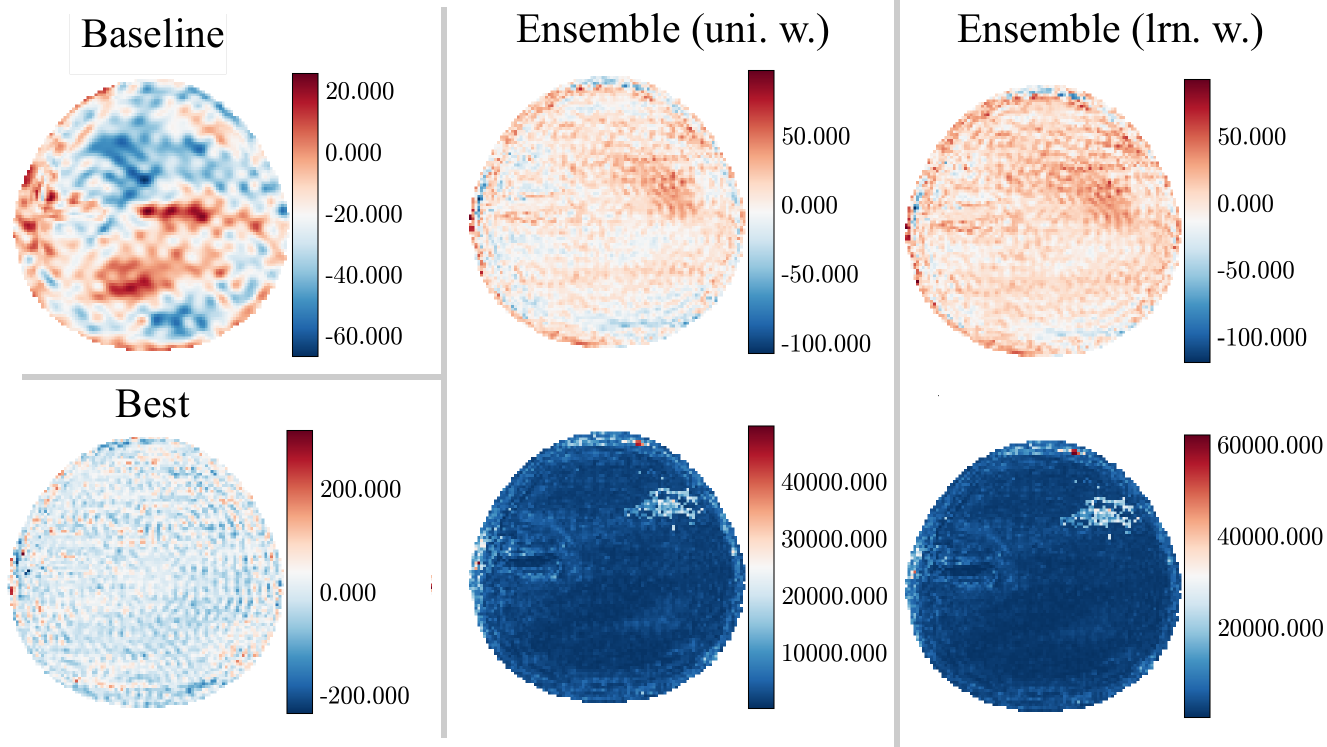}
    \captionof{figure}{Time averaged model estimated sensitivity of $J$ calculated using the
    temperature fields to $\kappa$. }
    \label{fig:sensitivity_compare}
\end{minipage}
\hfill
\begin{minipage}[c]{.48\textwidth}
    \centering
    \includegraphics[width=\linewidth]{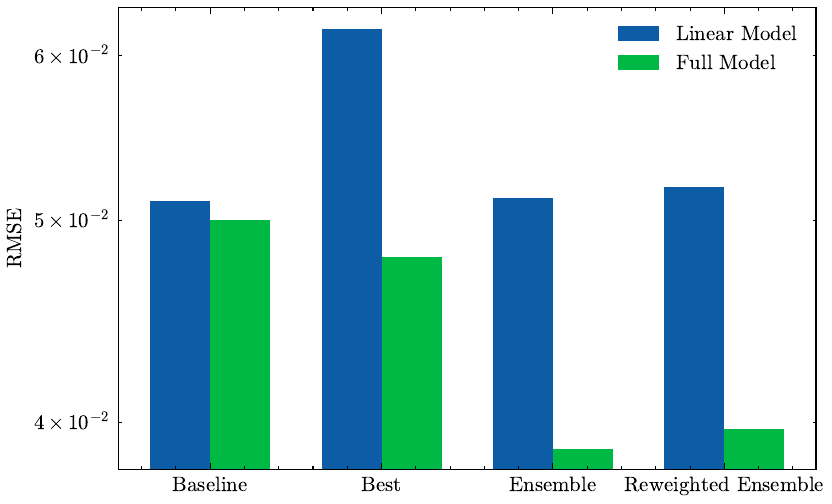}
    \captionof{figure}{RMSE of the temperature predictions from linearized models.}
    \label{fig:linear_model}
\end{minipage}
\vspace{1em}


\section{Conclusion}\label{sec:conclusion}

This paper presents a deep ensemble approach for improved parametric
sensitivity estimates for ocean models. Utilizing a large-scale HPO, we select
top performing models and use two weighting schemes to construct the ensembles.
The trained models are evaluated in single time-stepping prediction, long-range
autoregressive rollout, and parametric sensitivity estimation. Without
access to ground truth sensitivity, we evaluate the linearized models to assess
the quality of sensitivity estimates.  The results show that the proposed ensemble
models outperform the baseline model and best model from the HPO in all three
tasks and provide quantified uncertainty at the same time. Future work involves
extending the framework to multiple parameterizations and sensitivity evaluation 
through numerical differentiation of the physical model or data assimilation.

\begin{ack}
This research used resources from the NERSC, a U.S. Department of Energy Office
of Science User Facility located at LBNL. Material based upon work supported by
the U.S. Department of Energy, Office of Science, Office of Advanced Scientific
Computing Research and Office of BER, Scientific Discovery through Advanced
Computing (SciDAC) program, under Contract DE-AC02-06CH11357. We are grateful to
the Sustainable Horizons Institute's Sustainable Research Pathways workforce
development program.
\end{ack}

%
%
%
%
%
\bibliographystyle{unsrtnat}
\bibliography{ref}

\appendix
\newcommand{\hp}[1]{\texttt{#1}}
\section{Methodology details}
We list the detailed steps of the proposed framework in Algorithm~\ref{alg:1}. Note that 
this framework is not limited to top-K selection with uniform weighting or re-weighting via
linear regression. We plan to explore other model selection criteria, such as greedy selection~\cite{caruana2004ensemble}, in future work.
\begin{algorithm}
   \caption{Deep ensemble for ocean dynamics and parametric sensitivity} 
   \label{alg:1}
   \begin{algorithmic}[1]
       \STATE Given training and validation datasets, \(\mathcal{D}_{\text{train}}, \,  \mathcal{D}_{\text{val}}\) 
       \STATE Define baseline neural network, \(M_{\theta_{\text{base}}}\). 
       \STATE Select HPO surrogate \(S\) and acquisition function \(\alpha\).
       \STATE Select ensemble size \(K\).
       \STATE Initialize HPO with \(M_{\theta_{\text{base}}}, \, \theta^{\prime} \gets \theta_{\text{base}}\).
       \FOR{\(n\) in number of search steps}
       \STATE Evaluate candidate hyperparameters \(\theta^{\prime}\) using \(\mathcal{D}_{\text{val}}\).
       \STATE Fit \(S\) with \(\theta^{\prime}\) and compute \(\alpha(\theta^{\prime})\).
       \STATE Return new model candidates \(\theta^{\prime}\ = \arg \max_{\theta \in \chi} \alpha(\theta) \)
       \ENDFOR
       \STATE Rank models in HPO results and keep first \(K\) models.
       \STATE Fully train the selected models
       \IF{ Selection criterion is \texttt{Top-K}}
       \STATE Return model weights \(w_1 = \dots = w_K, \, \sum w_k=1\).
       \ELSIF{Selection criterion is \texttt{weighted}}
       \STATE Perform linear regression
       \STATE Return updated weights, \(w_1, \dots, w_K \).
       \ENDIF
       \STATE Produce ensemble predictions of \(\vect{x}_{i+1}\) and \( \frac{\p J}{\p \kappa}\).
   \end{algorithmic}
\end{algorithm}

\section{Experiment details}
\label{sec:experiment}

\subsection{Data generation}

Following \cite{sun2024parametric}, we perturb the Gent–McWilliams
parameterization while keeping all other parameterizations at their nominal
values to run independent simulations for model training and evaluation.
Specifically, we sample values from a uniform distribution (range in
Table~\ref{tab:parameterrange}) and create 100 forward runs. In each run, the
simulation is initialized from the same condition and integrated for 23 years,
with \textit{monthly} snapshots saved. At the 32 km resolution, the grid
consists of 8,521 hexagonal cells (a nearly circular horizontal domain), each
with 60 vertical levels. Each simulation year produces over 13 million ($8521
\times 60 \times 26$) cell values for each spatially and temporally varying
output variable in the dataset. We select five ocean states—\emph{layer
thickness, zonal velocity, meridional velocity, temperature}, and
\emph{salinity}—as targets and truncate the trajectories to retain the last 8
years, discarding the necessary spin-up stage. Each simulation run was performed
using the publicly available code, which can be found at
 \anonymize{https://github.com/anl-ecucuzzella/SOMAForwardCode}.

\begin{table}[h]
    \centering
    \caption{Range of Perturbed Parameter Values.}
    \begin{tabular}{cccc}
    \toprule 
    \textbf{SOMA Parameter} & \textbf{Symbol} &\textbf{Minimum} & \textbf{Maximum} \\
    \midrule
      GM\_constant\_kappa & $\kappa$ & 200.0 & 2000.0 \\
    \bottomrule
    \end{tabular}
    \label{tab:parameterrange}
\end{table}

The generated data, in its original fidelity and representation, poses
challenges for training FNO-based models because it is defined on an irregular
grid. To address this, the mesh-grid data was converted to a standard
latitude--longitude grid through spatial interpolation, and the values were
mapped to regular array entries. As a result, we obtain data on regular grids
stored as arrays, each instance having shape \((6, 60, 100, 100)\). The first
dimension contains the five ocean states and one model parameter \(\kappa\),
while the last three dimensions represent the spatial axes of the domain. We
then convert the last 8 years of trajectories from the 100 independent
simulations to these regular grid representations. Each simulation consists of
208 time steps covering the full trajectory. Finally, we split the trajectories
into training, validation, and testing sets and prepare input--output pairs
(Section~\ref{sec:method}) using two consecutive time steps.





\subsection{Data preprocessing}

The raw data are three-dimensional in space, forming a nearly circular horizontal
domain with multiple vertical layers indicating depth. Similar to most
weather and climate modeling tasks~\cite{price_gencast_2023, nguyen2023climax}, we
treat the vertical layers as additional feature dimensions. The data are therefore
represented with shape \((360, 100, 100)\), where the first axis corresponds to
various ocean state values across depths. For example, the first 60 entries store
layer thickness values from the sea surface (depth 0) to the 60th layer, while
the last 60 represent the parameterization for this particular simulation across
the vertical layers, which remain spatially constant in this work.

We transform the data for model training to improve stability and
generalization. For each ocean state at each horizontal location (including the model output $\vect{y} = \vect{x_{t+1}}$), we use
\emph{per-depth} statistics to standardize the data as follows, for all \(t\),
\[
\vect{x}_{i, d}^{\text{scale}} = \frac{\vect{x}_{i,d} - \mu_{i,d}}{\sigma_{i,d}},
\]
where \(\vect{x}_{i, d}\) denotes the \(i\)th ocean state at depth \(d\), and
\(\mu_{i,d}\) and \(\sigma_{i,d}\) are the corresponding sample mean and standard
deviation obtained from the training set. At evaluation time, after the trained
model produces forecasts, we apply inverse scaling to map the values back to the
original physical space, for all \(t\),
\[
\vect{x}_{i,d} = \vect{x}_{i, d}^{\text{scale}} \cdot \sigma_{i,d} + \mu_{i,d}.
\]

For the variance of predictions from the ensemble model, the corresponding value
in the original space can be obtained as
\[
\mathrm{Var}[\vect{x}_{i,d}] = \sigma^2_{i,d} \cdot \mathrm{Var}[\vect{x}_{i, d}^{\text{scale}}].
\]

We leverage automatic differentiation to compute the parametric sensitivity.
Since the trained networks operate on normalized data and \(\frac{\partial
J(\vect{x}_{t+1})}{\partial \kappa}\) is not computable from \(\frac{\partial J
(\vect{x}_{t+1}^{\text{scale}})}{\partial \kappa^{\text{scale}}}\) without the
explicit access to \(\frac{\partial \vect{x}_{t+1}^{\text{scale}}}{\partial
\kappa^{\text{scale}}}\) (explanation below), we wrap the network operation with additional normalization and its inverse
transform and differentiate it with respect to the parameterization \(\kappa\).
Therefore, the output directly reflects the estimated \(\frac{\partial
J(\vect{x}_{t+1})}{\partial \kappa}\).

\paragraph{Rescaling computed sensitivity} \label{app:unnorm_sensitivity}
Given that we train the neural networks using normalized data, the most
straightforward way to compute the adjoint sensitivity is to pass the model
output to the scalar-value objective function and differentiate this function
with respect to the input parameterization, \(\frac{\p J^{\scale}}{\p
\kappa^{\scale}}\), and perform associated transformations to reach \(\frac{\p
J}{\p \kappa}\). However, we now show that this is not feasible without explicit
access to \( \frac{\p \vect{x}_{t+1}^{\scale}}{\p \kappa^{\scale}} \). For
simplicity, we now use \(\vect{x}\) to denote the model output
\(\vect{x}_{t+1}\).

With \(J(\vect{x}) = \frac{1}{2} \Vert \vect{x} \Vert_2^2 \) and normalization
transformation listed in Section~\ref{sec:exp},  in the normalized space, we can easily 
compute

\begin{equation}\label{eqn:6}
    \frac{\p J(\vect{x}^{\scale})}{\p \kappa^{\scale}} =   (\vect{x}^{\scale})^{\top} \frac{\p \vect{x}^{\scale}}{\p \kappa^{\scale}}.
\end{equation}

Our final objective is to compute the sensitivity in the original scale, plugging (\ref{eqn:6}) in,

\begin{equation}\label{eqn:7}
\begin{aligned}
    \frac{\p J(\vect{x})}{\p \kappa} &=   \vect{x}^{\top} \frac{\p \vect{x}}{\p \kappa} = \vect{x}^{\top} \frac{\sigma_{\vect{x}}}{\sigma_{\kappa}} \frac{\p \vect{x}^{\scale}}{\p \kappa^{\scale}}  \\
    &= \frac{ (\sigma_{\textbf{x}}\vect{x}^{\scale} + \mu_{\vect{x}})^{\top} \frac{\sigma_{\vect{x}}}{\sigma_{\kappa}} \frac{\p \vect{x}^{\scale}}{\p \kappa^{\scale}} \frac{\p J(\vect{x}^{\scale})}{\kappa^{\scale}} } {(\vect{x}^{\scale})^{\top} \frac{\p \vect{x}^{\scale}}{\p \kappa^{\scale}}}
    .
\end{aligned}
\end{equation}

Since \( \mu_{\vect{x}} \neq 0 \) and \(\vect{x}\) is not a scalar, we require
the access to \( \frac{\p \vect{x}^{\scale}}{\p \kappa^{\scale}} \) to perform
the transformation in (\ref{eqn:7}), which requires additional and less
efficient steps to obtain in the reverse-mode auto-differentiation process.
Therefore, we use the objective function directly taking the input in the
original scale and add additional normalizing and unnormalizing steps to
directly compute \( \frac{\p J(\vect{x})}{\p \kappa} \).

\subsection{Model training and evaluation}
We adopt the FNOs implemented in \texttt{PhysicsNeMo}~\cite{physicsnemo2023} and use the default hyperparameters as the baseline for all surrogates
in this work.
The domain of the ocean states is nearly circular in the horizontal direction,
and the horizontal mask is not constant across vertical layers because the
domain of interest is bowl-shaped, with horizontal area decreasing as depth
increases. As a result, we apply masking during training and compute the loss
only over values inside the domain. No special constraints are placed outside
the domain; both true and predicted values are simply set to zero.

Once the models are trained, we evaluate them on the testing set and report the
root mean squared error (RMSE) of model performance across various state
variables for single-step forward predictions. In addition, we use the
\emph{time-averaged} ocean states as a constant predictor to set a benchmark for
model evaluation. Different from training, we do not set values outside the
domain to zero; instead, we exclude them from the calculation and only account
for values within the domain of interest. 

We randomly select a simulation associated with a unique \(\kappa\) in the
testing set to evaluate and visualize model performance on autoregressive
rollout and adjoint sensitivity estimation. For the rollout, we compute the
spatially averaged values of each ocean state at each time step and compare them
to the ground truth. For adjoint sensitivity, since true derivatives are not
available, we report only the time-averaged adjoint sensitivities of variables
at different vertical levels.

\subsection{Hyperparameter optimization details}

\begin{figure}[H]
    \centering
    \includegraphics[width=.6\linewidth]{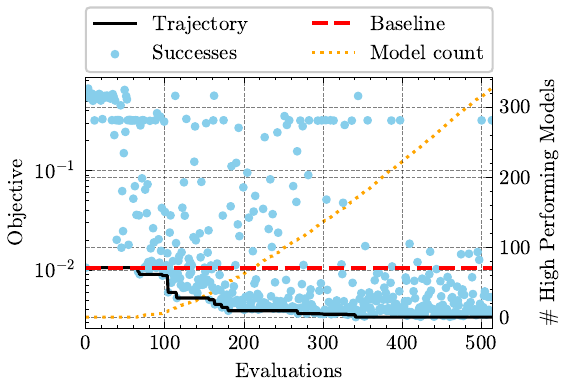}
    \caption{Hyperparameter search trajectory. Starting with the baseline configuration, 
    the search balances the exploration and exploitation which leads to increasingly better models
    depicted by the solid black line. The number of high performing models increases accordingly.}
    \label{fig:hpo_traj}
\end{figure}

With the baseline model, we aim to conduct large-scale HPO to (1) further
improve predictive accuracy and (2) create a candidate pool for constructing a
deep ensemble for stable rollout and improved adjoint sensitivity estimation
with quantified uncertainty. We utilize a \emph{12-dimensional} hyperparameter
search space spanning FNO architectures, data transformations, and training
strategies. A detailed description of the search space is provided in
Table~\ref{tab:hpo}. The optimization objective is set to minimizing the
validation loss, and we adopt the 1-epoch strategy~\cite{egeleOneEpochAll2023}
to avoid excessive computational overhead during the search. We employ 40 NVIDIA
A100 GPUs for parallelized evaluations, obtaining over 500 hyperparameter
configurations. Figure~\ref{fig:hpo_traj} shows the search trajectory, which
gravitates toward hyperparameter configurations outperforming the baseline
(marked in red). This is accompanied by an increased number of high-performing
models compared to the baseline. We then construct ensembles from the top-$K$
configurations and train each member for 64K steps. Finally, we apply linear
regression to re-weight the ensemble members, as discussed in
Section~\ref{sec:exp}, to further improve performance.

\begin{table}[h]
    \centering
    \caption{Hyperparameter search space}\label{tab:hpo}
    \begin{tabular}{p{3cm}p{1cm}p{5cm}p{3cm}}
    \toprule
        Variable Names.      & Type & Range/Choice                                   & Baseline\\
    \midrule
        \hp{padding}         & \hp{int} & \([1, 16]\)                                  &  8 \\
        \hp{padding\_type}   & \hp{str}  & \hp{[`constant', `reflect', `replicate', `circular']}                      & \hp{constant}\\
        
        \hp{coord\_feat}     & \hp{bool} & \hp{[True, False]}                                  & \hp{True} \\
        \hp{lift\_act} & \hp{str} & \hp{[`relu', `leaky\_relu', `prelu', `relu6', `elu', `selu', `silu', `gelu', `sigmoid', `logsigmoid', `softplus', `softshrink', `softsign', `tanh', `tanhshrink', `threshold', `hardtanh', `identity', `squareplus']}  & \hp{gelu}\\
        \hp{num\_FNO} & \hp{int} & \([2, 32]\) & 4\\
        \hp{num\_modes} & \hp{int} & \([2, 32]\) & 16\\
        \hp{latent\_ch} & \hp{int} & \([2, 64] \) & 32\\
        \hp{num\_projs} & \hp{int} & \([1, 16]\) & 1 \\
        \hp{proj\_size} & \hp{int} & \([2, 32]\) & 32\\
        \hp{proj\_act} & \hp{str} & \hp{[`relu', `leaky\_relu', `prelu', `relu6', `elu', `selu', `silu', `gelu', `sigmoid', `logsigmoid', `softplus', `softshrink', `softsign', `tanh', `tanhshrink', `threshold', `hardtanh', `identity', `squareplus']} & \hp{silu}\\
        \hp{lr} & \hp{float} & \([10^{-6}, 10^{-2}]\) & \( 10^{-3} \)\\
        \hp{weight\_decay} & \hp{float} & \([0, 0.1]\) & 0\\
    \bottomrule
    \end{tabular}
\end{table}

\subsection{Additional results}
We list the visualizations of the single time-stepping prediction,
autoregressive rollout, and parametric sensitivity estimates of all ocean states
in this section.
\begin{figure}[!htpb]
    \centering
    \includegraphics[width=0.8\linewidth]{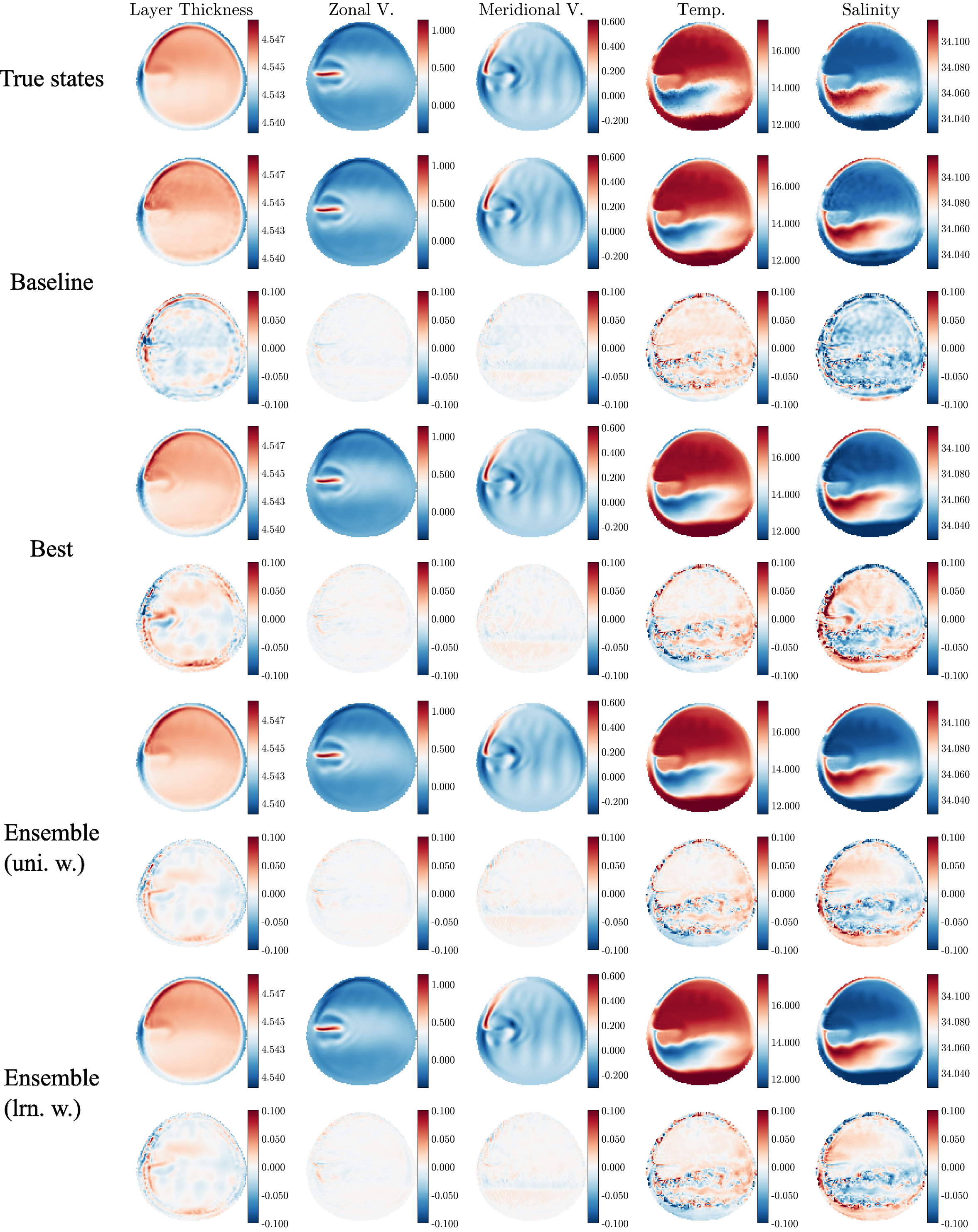}
     \caption{Visualization of the single-step model predictions at the sea
     surface. The top most row shows the true ocean states; in each inset
     thereafter, the top row is the predicted states and bottom row shows the
     fractional error of the predicted fields.}
    \label{fig:single_all_states}
\end{figure}

For the single time-stepping prediction of all ocean states,
Figure~\ref{fig:single_all_states} visualizes the baseline performance on a
testing data point. At the sea surface (depth 0), the baseline model captures
state fields that closely resemble the true ocean states. However, compared to
the zonal and meridional velocities, forecasts of the other states exhibit
relatively larger fraction errors. Such errors can reduce the reliability of
adjoint sensitivity estimates and accelerate error accumulation during
autoregressive rollout, underscoring the need for improved modeling of ocean
dynamics. For this particular case, the best model from HPO does not show a
clear improvement over the baseline, reflecting only marginal performance
gains. In contrast, the Top-10 ensemble improves the forecast of layer
thickness, yielding lower fraction errors than both the baseline and optimal
models. The weighted Top-10 ensemble performs similarly to the uniform Top-10
ensemble, with both ensembles outperforming the baseline and single best
models.

\begin{figure}[h]
\centering
\begin{subfigure}{.8\linewidth}
    \centering
    \includegraphics[width=\linewidth]{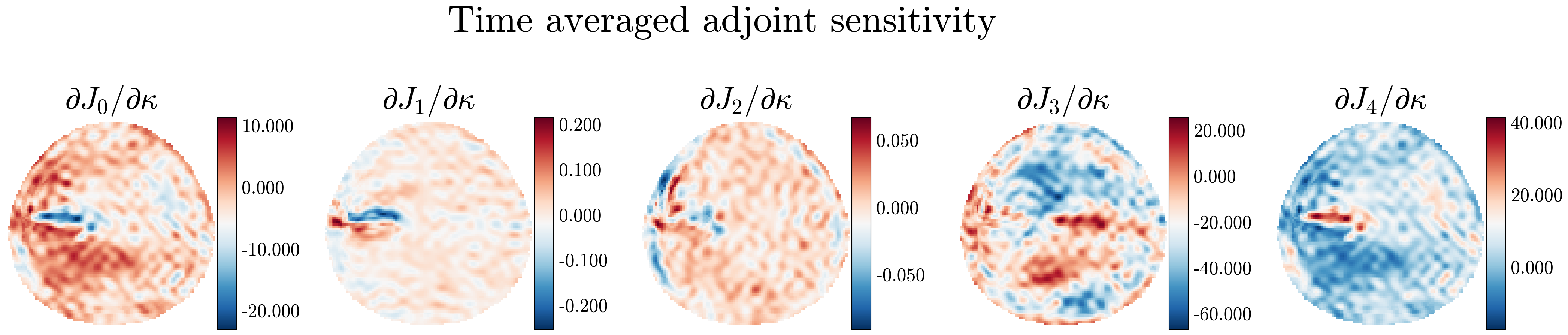}
    \caption{}\label{fig:baseline_adjoint}
    \vspace{2em}
\end{subfigure}

\begin{subfigure}{.8\linewidth}
    \centering
    \includegraphics[width=\linewidth]{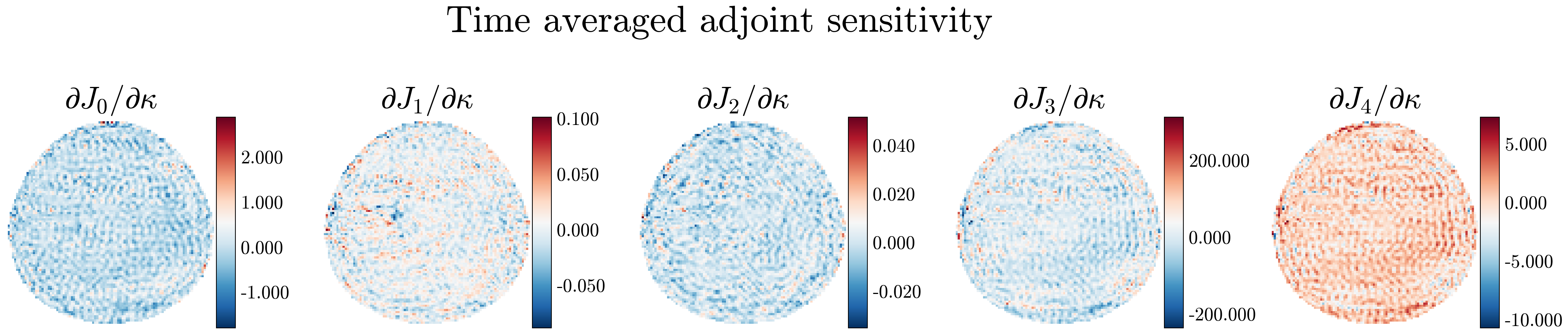}
    \caption{}\label{fig:optimal_adjoint}
    \vspace{2em}
\end{subfigure}

\begin{subfigure}{.8\linewidth}
    \centering
    \includegraphics[width=\linewidth]{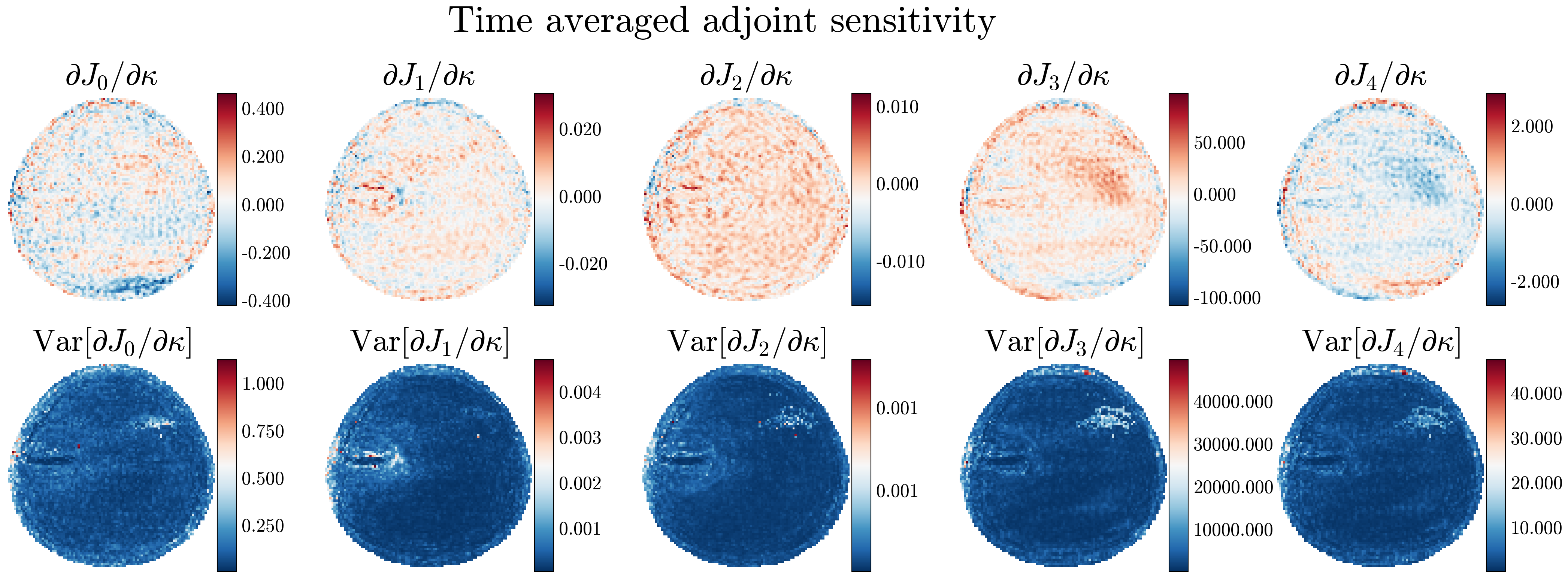}
    \caption{}\label{fig:top10_adjoint}
    \vspace{2em}
\end{subfigure}

\begin{subfigure}{.8\linewidth}
    \centering
    \includegraphics[width=\linewidth]{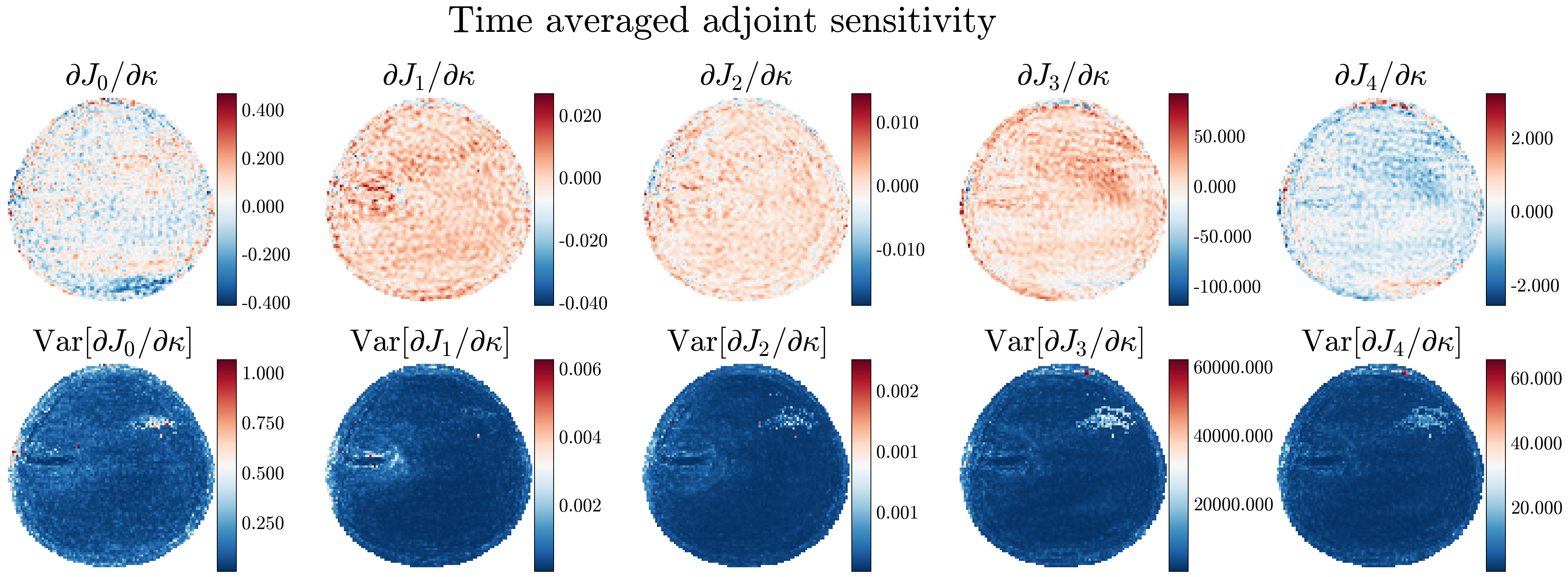}
    \caption{}\label{fig:wtop10_adjoint}
    \vspace{2em}
\end{subfigure}
    \caption{Time-averaged parametric sensitivity of ocean states at
    the sea surface with respect to the parameterization. The order of the states are
    layer thickness, zonal velocity, meridional velocity, temperature, and salinity. (a) baseline; (b) optimal model; (c) Top-10 ensemble; (d) Weighted Top-10 ensemble.}
    \label{fig:adjoint}
\end{figure}

Figure~\ref{fig:adjoint} shows the parametric sensitivity estimates from all
models considered in this work across the ocean states. Although the baseline
and best models produce similar forward predictions, their sensitivity estimates
differ substantially. In contrast, the two ensembles exhibit only marginal
differences in sensitivity estimates and show strong agreement in predictive
uncertainty.

\end{document}